\documentclass[%
 aip,
 apl,
 amsmath,amssymb,
 reprint,%
]{revtex4-1}

\usepackage{graphicx}
\usepackage{dcolumn}
\usepackage{bm}

\usepackage[utf8]{inputenc}
\usepackage[T1]{fontenc}
\usepackage{mathptmx}
\begin{document}

\preprint{AIP/123-QED}

\title{Position Sensitive Response of a Single-Pixel Large-Area SNSPD}%

\author{Claire E. Marvinney}
\email{marvinneyce@ornl.gov}
\affiliation{Oak Ridge National Laboratory, 1 Bethel Valley Rd, Oak Ridge, TN 37831}
\author{Brian E. Lerner}
\affiliation{Oak Ridge National Laboratory, 1 Bethel Valley Rd, Oak Ridge, TN 37831}
\author{Alexander A. Puretzky}
\affiliation{Oak Ridge National Laboratory, 1 Bethel Valley Rd, Oak Ridge, TN 37831}
\author{Aaron J. Miller}
\affiliation{Quantum Opus LLC, 22241 Roethel Dr Ste A, Novi, MI 48375}
\author{Benjamin J. Lawrie}
\email{lawriebj@ornl.gov}
\affiliation{Oak Ridge National Laboratory, 1 Bethel Valley Rd, Oak Ridge, TN 37831} 
\footnote{This manuscript has been authored by UT-Battelle, LLC under Contract No. DE-AC05-00OR22725 with the U.S. Department of Energy. The United States Government retains and the publisher, by accepting the article for publication, acknowledges that the United States Government retains a non-exclusive, paid-up, irrevocable, world-wide license to publish or reproduce the published form of this manuscript, or allow others to do so, for United States Government purposes.  The Department of Energy will provide public access to these results of federally sponsored research in accordance with the DOE Public Access Plan (http://energy.gov/downloads/doe-public-access-plan).}
\date{\today}

\begin{abstract}
Superconducting nanowire single photon detectors (SNSPDs) are typically used as single-mode-fiber-coupled single-pixel detectors, but large area detectors are increasingly critical for applications ranging from microscopy to free-space quantum communications.  Here, we explore changes in the rising edge of the readout pulse for large-area SNSPDs as a function of the bias current, optical spot size on the detector, and number of photons per pulse.  We observe a bimodal distribution of rise times and show that the probability of a slow rise time increases in the limit of large spot sizes and small photon number. In the limit of low bias currents, the dark-count readout pulse is most similar to the combined large spot size and small-photon-number bright-count readout pulse.  These results are consistent with a simple model of traveling microwave modes excited by single photons incident at varying positions along the length of the nanowire.

\end{abstract}

\maketitle

Superconducting nanowire single photon detectors (SNSPDs) have become increasingly relevant to quantum networking \cite{hadfield2006quantum,takemoto2015quantum,zhang2018experimental,sun2016quantum,graffitti2020measurement} and quantum sensing \cite{hochberg2019detecting,ahmed2018quantum} because they offer high-speed, high-quantum-efficiency, and low-dark-count-rate single photon detection \cite{natarajan2012superconducting,eisaman2011invited}. Readout pulse analysis has been used to provide spatially resolved detection in a single channel SNSPD \cite{zhao2017imager}, model geometric variations in device design \cite{zhao2018distributed}, provide few-photon number resolution~\cite{cahall2017multi,nicolich2019universal, zhu2020resolving}, and design impedance matched circuits\cite{zhu2020resolving} for large-area SNSPDs \cite{zhang2019nbn}. In parallel with these device developments, a growing demand has emerged for large-area SNSPDs capable of providing high collection efficiency for applications ranging from microscopy to satellite communications \cite{zhao2017imager,li2015large,wang2019large, allmaras2020demonstration}. However, the signals produced by large-area SNSPDs are not always directly comparable to those produced by single-mode SNSPDs. A fundamental understanding of the waveforms produced by large-area SNSPDs is therefore essential to future progress in these fields.

In current large-area SNSPDs, the kinetic inductance (proportional to the nanowire length) can be over an order of magnitude larger than that of a single-mode SNSPD \cite{zhang2019nbn}, leading to poor impedance matching using traditional electronics and thus a decreased maximal count rate \cite{lv2017large}, and increased detector reset time \cite{clem2012kinetic}.  The increased nanowire length also increases the signal propagation delay. A 2 mm long nanowire has a propagation delay on the order of the hot-spot formation time of ~200 ps \cite{zhao2018distributed, berggren2018superconducting}.  Together, the increased kinetic inductance and propagation delay of large-area SNSPDs must be considered when designing these devices and characterizing and modeling the waveforms of the microwave modes in their long meander lines \cite{zhao2018distributed}.

\begin{figure}[t!]
    \centering
    \includegraphics[width=\columnwidth]{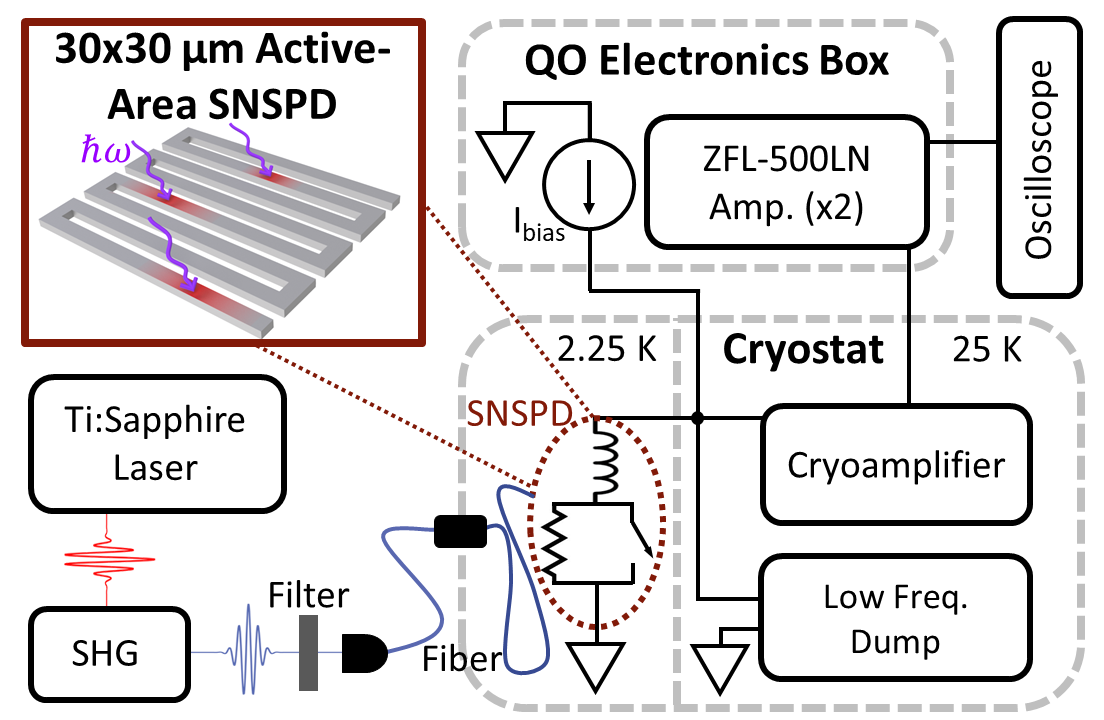}
    \caption{Schematic of the experiment used for collecting rise time histograms and waveforms of the SNSPD readout pulse, using the typical lumped element depiction of an SNSPD hotspot. Inset of the superconducting nanowire meander line during hotspot formation. }
    \label{fig:schematic}
\end{figure}

Here, we use high-speed signal analysis to characterize the waveforms generated by large-active-area (30$\mathrm{\mu m}$ x 30$\mathrm{\mu m}$) SNSPDs optimized for visible photon detection and integrated with low-noise cryogenic and room temperature amplifiers.  A general schematic of the experiment is shown in Fig.~\ref{fig:schematic}. The rising edge of the readout pulse is characterized as a function of photon number per pulse ($\mu$), bias current, and spot size on the detector. SNSPDs are not generally used for $\mu>1$ because they do not typically provide photon number resolution. Similarly, they are typically operated at a bias current that produces of order 1-10 dark counts per second. However, characterizing the changes in readout waveforms for large $\mu$ and large bias currents helps to generate a basic understanding of the  waveforms acquired for typical single-photon counting operating parameters.  Additionally, varying the spot size on the detector allows us to explore spatial variations in the readout waveforms that do not appear in smaller single-mode detectors.  

\begin{figure}[b]
    \centering
    \includegraphics[width=\columnwidth]{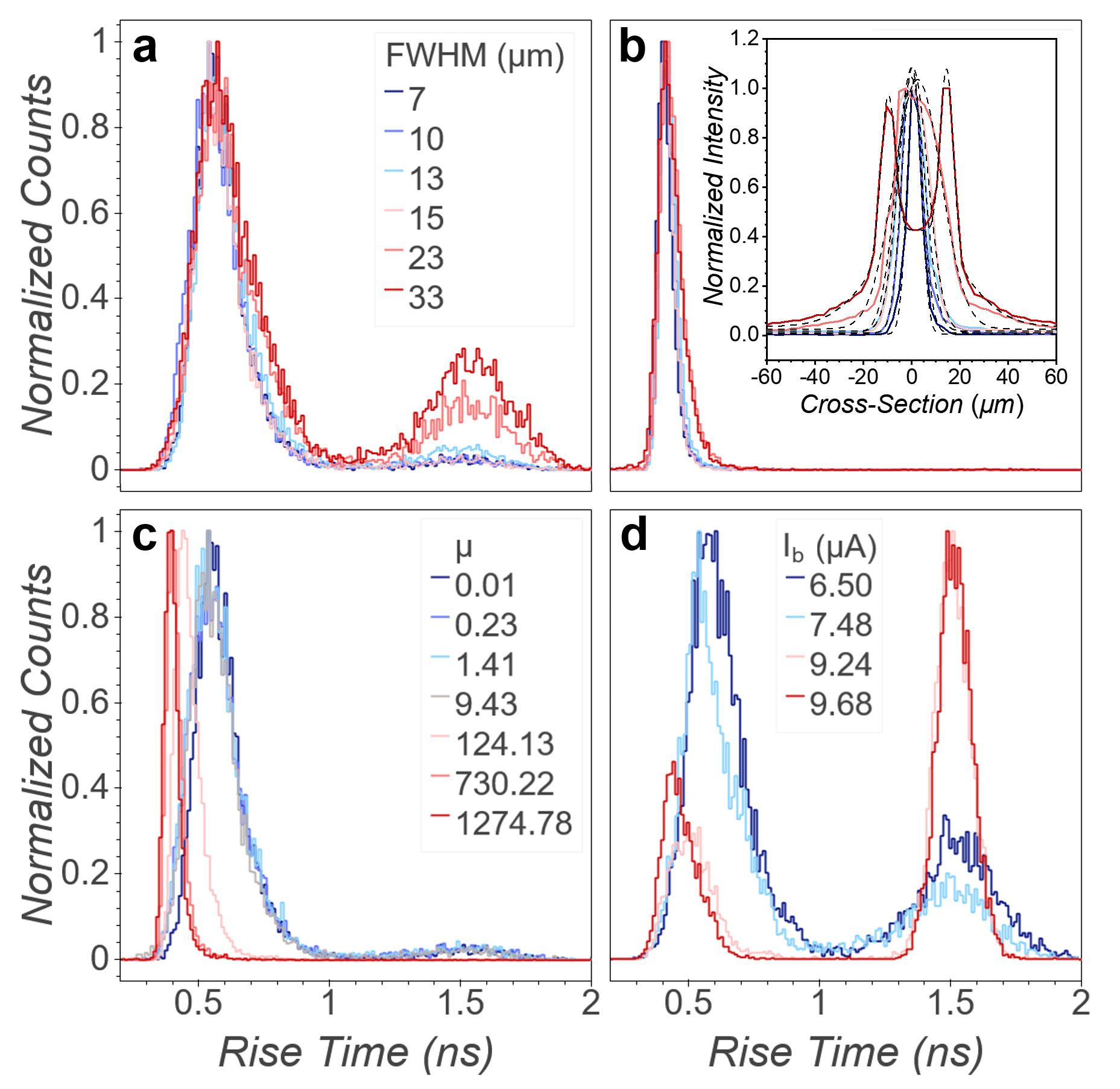}
    \caption{SNSPD rise-time histograms with mean photon number per pulse $\mu$ = 0.01 (a) and $\mu$ = 700 (b) for varying spot size on the detector (legend in (a)). Cross-sections of the beam profiles at the tip of each fiber are shown in the inset of (b) with corresponding fits. The same multimode fiber was used to generate ~23$\mu m$ and ~33$\mu m$ spots, with modified fiber incoupling resulting in a shift from a quasi-Gaussian mode to a donut mode. (c) Bright-count rise-time histograms for variable $\mu$ and a 7 $\mu m$ spot size. (d) Intrinsic dark-count rise-time histograms for variable bias current.}
    \label{fig:histograms}
\end{figure}

The detectors were cooled to 2.25K and held at a constant bias current of 6.50$\mathrm{\mu A}$, where the detectors have a dark-count rate of 2-4/s.   Dark-count waveforms were subsequently acquired for bias currents of 6.50$\mathrm{\mu A}$ to 9.68$\mathrm{\mu A}$, corresponding to dark-count rates spanning six orders of magnitude. The attenuated second harmonic of a Ti:sapphire oscillator with a 2 ps pulse duration, 400 nm wavelength, and 1 MHz rep rate was used as the light source for all experiments.  Mean photon number per pulse was varied from $\mu = 0.01$ to $\mu = 700$. The light was fiber coupled from the laser using a variety of fiber patch cables in order to change the spot size  of the light on the SNSPD, as shown in the inset of Fig.~\ref{fig:histograms}b. Readout pulse signals were collected on a 33 GHz oscilloscope. The resulting waveforms and histograms of the waveform rise time are presented here.
  
As the optical spot size was varied, no notable changes were observed in the signal amplitude, so we focus here on the rise-time distribution.  For small mean photon number, a bimodal distribution exists in the histogram of rise times as shown in Fig.~\ref{fig:histograms}a, with the centroids of the two modes at 0.57 ns and 1.50 ns and a 2x greater probability of measuring a rise-time that falls in the faster mode. Larger spot sizes substantially increase the probability of a slower rise time for a constant photon count rate.  When illuminated with a 33 $\mu$m donut mode, the rise time is 6 times more likely to fall in the slower mode than when illuminated with a 7 $\mu$m spot size.  However, for average photon number much larger than 1, the slow mode of the bimodal distribution disappears, and the centroid of the faster mode moves to 0.41 ns for all beam spot sizes, as shown in Fig.~\ref{fig:histograms}b. 

The relationship between rise time and $\mu$ becomes more apparent in the histogram of rise times for constant $\mathrm{7 \mu m}$ spot size and variable $\mu$ shown in Fig.~\ref{fig:histograms}c. Only the fast component of the readout waveform is present for this relatively small spot size, and for $\mu <10$, the rise-time distribution is independent of $\mu$.  As $\mu$ continues to increase  to 1275, the the centroid of the rise-time distribution decreases from 0.57 ns to 0.40 ns and with a corresponding reduction in the full-width half-maximum (FWHM) of the distribution from 0.18 ns to 0.06 ns.

Figure~\ref{fig:histograms}d illustrates the intrinsic dark-count rise-time histograms for variable bias current. For bias currents of 6.5 $\mathrm{\mu A}$ and 7.48 $\mathrm{\mu A}$ (corresponding to dark-count rates of 2/s and 100/s respectively), the bimodal distribution with centroids of 0.59 ns and 1.53 ns and a ratio between the faster and slower modes of 2 to 1 is consistent with the bright-count rise-time histograms for small $\mu$ and large spot size in Fig.~\ref{fig:histograms}a. However, as the bias current increases to 9.24 and 9.68 $\mathrm{\mu A}$ (corresponding to dark count rates of 100,000/s and 3,000,000/s), the device moved into an electrothermal oscillation regime in which a hostpot periodically forms and quenches with frequency determined by the device inductance and load impedance~\cite{hadfield2005low,kerman2009electrothermal,liu2013electrical,berggren2018superconducting}.  In this high bias current limit, the relative probability of the slow histogram mode climbs to 2x that of the fast mode, and the faster of the dark-count rise-time peaks shifts from 0.59 ns to 0.46 ns, consistent with the change in the fast bright-count rise-time distribution with increasing $\mu$ shown in Fig.~\ref{fig:histograms}c, though the FWHM of the faster peak does not decrease as substantially with bias current.

\begin{figure}[t]
    \centering
    \includegraphics[width=\columnwidth]{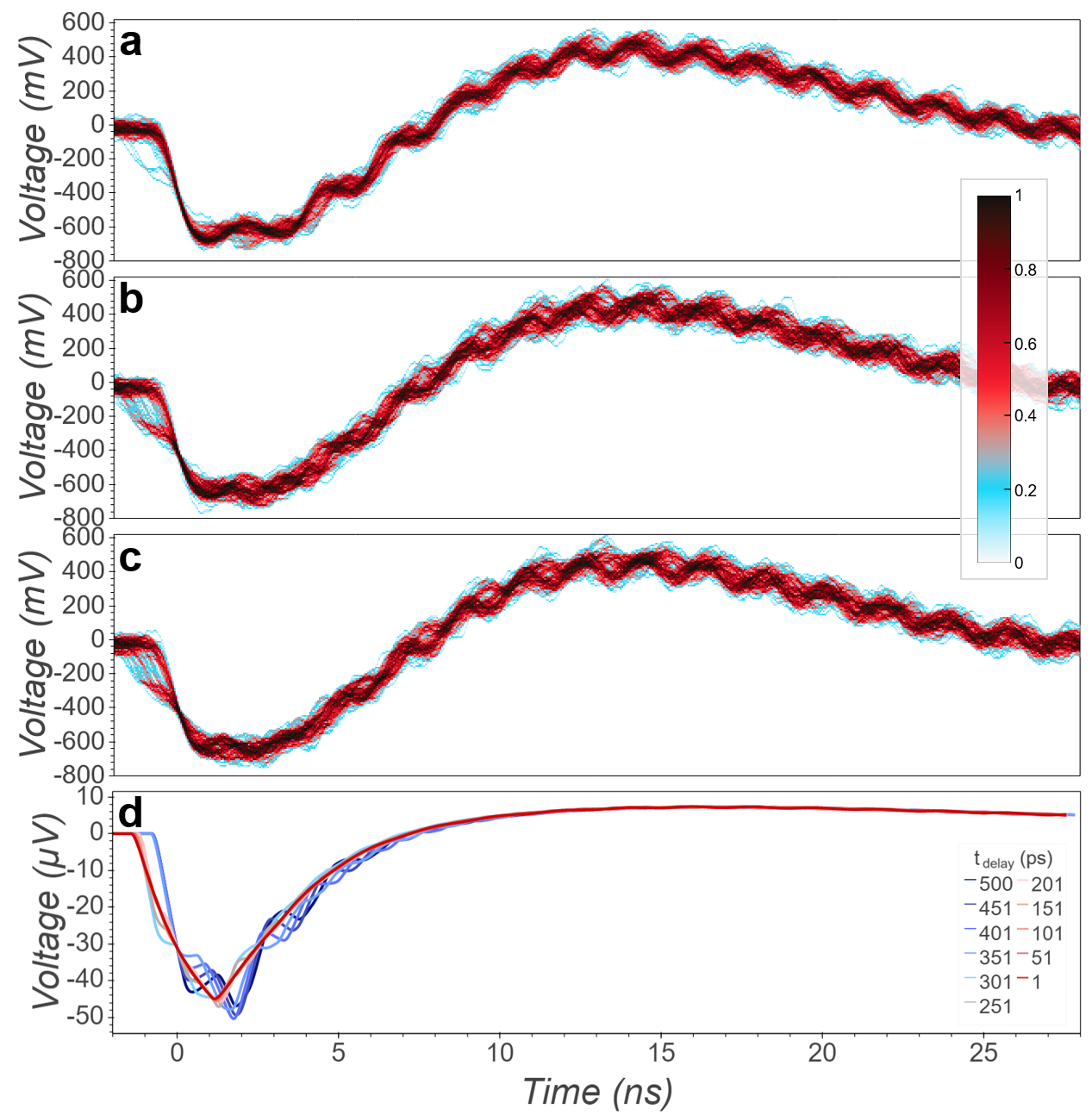}
    \caption{100 waveforms of the bright count data collected in digital persistence mode at $\mu$ = 0.01 and for spot sizes of (a) 7 $\mu$m and (b) 23 $\mu$m. (c) 100 waveforms of dark count data for the typical operating bias current of 6.50 $\mu$A. Waveforms are inverted by the cryoamplifier. (d) Simulated waveforms for photons incident at 11 positions on the nanowire defined by the microwave propagation time along the nanowire (shorter propagation times correspond to positions closer to the device readout). }
    \label{fig:waveforms}
\end{figure}

While there are correlations evident between the histograms plotted in Fig.~\ref{fig:histograms}, it is difficult to fully understand these correlations without examining the SNSPD readout waveforms directly. Figures~\ref{fig:waveforms}a and b depict 100 waveforms in a digital persistence mode acquired for 7 $\mu$m and 23 $\mu$m spot sizes and $\mu$ = 0.01. Two notable characteristics of the waveforms are (1) the 0.4-0.6 GHz sinusoidal modulation of the signal, and (2) the overshoot past 0~V on average at 8.45 ns after the initial pulse.  The sinusoidal modulation can substantially increase the rise time, an effect that occurs more frequently for larger spot sizes.

An LTspice model \cite{LTspice} was created to predict whether variation in the photon arrival position on the SNSPD could lead to characteristics (1) and (2), using a lumped element depiction of the hotspot as shown in Fig.~\ref{fig:schematic}. The simulated position-dependent waveforms are shown in Figure~\ref{fig:waveforms}d. A kinetic inductance of 10 $\mu$H and square resistance of 575 $\Omega$  is assumed based on experimental measurements of smaller devices.  The 5 mm total nanowire length yields a microwave delay time of approximately 500 ps, based on the approximate speed of light in the nanowire of $v = 0.02c$ \cite{zhao2018distributed, berggren2018superconducting}.  Unlike the lumped element model of an SNSPD, the position of a photon incident on the nanowire is thus simulated by varying the microwave propagation time after initial photon absorption, with shorter time delays closer to the readout circuit. The pulse is simulated using a 2 k$\Omega$ local resistance-per-square with a 2 ns on-time voltage switch representing the slower hotspot formation across multiple squares in the high impedance nanowires.  The sinusoidal oscillations of characteristic (1) are found to be 0.50-0.70 GHz in the simulation, in rough agreement with the experiment frequency and also amplitude when the impedance of the the nanowire meander line is 300 $\Omega$. In practice, the SNSPDs were connected to a high-pass filter before going through the amplification circuit, with a 3dB high-pass knee at 33MHz and a 3dB low-pass knee at 1GHz. The filter and poor impedance matching of the amplifiers with the meander line lead to the characteristic (2) overshoot of 0V ranging from 7.89-8.71 ns, in agreement with experiment.  The simulated waveform is a representation of the raw SNSPD pulse before amplification and is consistent with a combined cryo- and room-temperature- gain of approximately 12,000.

The frequency and amplitude of the oscillations in the simulated SNSPD readout pulse vary as the incident photon position is varied from the grounded end (500 ps) to the coax connected end (1 ps) of the nanowire delay line as shown in Fig.~\ref{fig:waveforms}d.  At the grounded end of the nanowire, a  20-80 percent rise time of 0.60 ns is observed; toward the  middle of the nanowire, the oscillation frequency increases, and the rise time begins to slow as a forms in the rising edge; toward the coax connected end of the of the nanowire, the oscillations decrease in amplitude and form multiple ripples in the rising edge of the signal, leading to slower rise times of 1.44 ns. These modeled variations in the SNSPD rise time are qualitatively consistent with the measured bimodal rise-time histograms for large spot sizes that illuminate the entirety of the nanowire meander line.  For the largest spot sizes, photons incident near the front of the detector yield slower rise times, while faster rise times are measured for photons incident toward the center and back of the nanowire. The slow rise times are negligible for the smaller spot sizes  in Fig.~\ref{fig:histograms}a that only illuminate the center of the device. The simulation corroborates this; photons incident near the center and back of the nanowire delay line are the fastest.

The meander line position dependence evident in the experimental and modeled data can thus be understood as a result of the roughly 500 ps propagation delay in the 5 mm long meander line and the corresponding pulse echo that generates the 0.4-0.6 GHz sinusoidal modulation of the readout pulse.  Treating the impedance mismatched meander line as a shorted microwave line, an oscillation frequency of 0.5 GHz in a 5 mm meander line would lead to a microwave propagation speed of 0.017c, consistent with previous works \cite{zhao2018distributed, berggren2018superconducting}.  At the backside of the nanowire, the microwave signal can only propagate in a single direction, but as the detection event approaches the front side of the nanowire, the signal can also propagate backwards, leading to pulse echo interference with the forward propagating signal. This effect results in the bimodal rise-time histogram for small $\mu$, large area pulses seen in Fig.~\ref{fig:histograms}a.  Similarly, the bimodal dark-count rise-time histogram for small bias currents in Fig.~\ref{fig:histograms}d can be understood as a result of roughly uniformly distributed dark count events across the length of the meander line. The dark-count waveforms at low bias currents, shown in Fig.~\ref{fig:waveforms}c, appeared nearly identical to the large spot-size bright count waveforms. Thus, under typical operating conditions, dark counts are not triggered from a single defect in the nanowire meander line, but instead from a random selection from a large distribution of defects across the entire span of the detector, consistent with a large spot size photon source and other recent experiments~\cite{zhao2017imager}.  

\begin{figure}[t]
    \centering
    \includegraphics[width=\columnwidth]{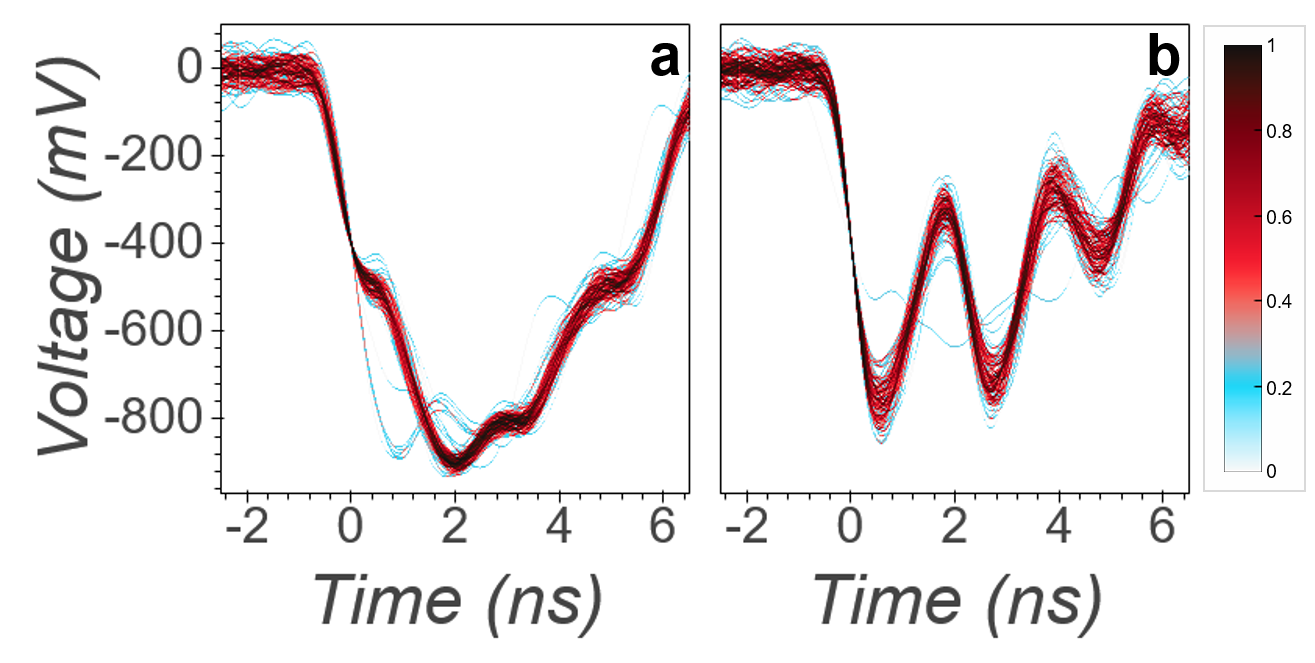}
    \caption{100 waveforms collected in digital persistence mode under dark conditions at the highest bias current of 9.68 $\mu$A where electrothermal fluctuations lead to 3 Mcps (a) and under bright conditions with a high $\mu$ = 700, typical bias current of 6.50 $\mu$A, and a 33 $\mu$m spot size sending photons to the entire detector (b) (waveforms are inverted by the cryoamplifier). }
    \label{fig:highcases}
\end{figure}

While the LTspice simulation describes the position dependence of the signal readout pulse under typical single-photon counting operating conditions, it does not yet describe the limiting cases of high bias currents when the device is exhibiting electrothermal oscillations or of high $\mu$ where locations across the entire device are being driven normal.  In the high bias current electrothermal oscillation regime, both modes of the bimodal rise time distribution have smaller FWHMs (bias currents > 7.48 $\mu$A in Fig.~\ref{fig:histograms}d) than under typical operating conditions and the waveforms in Fig.~\ref{fig:highcases}a have a uniform appearance. This demonstrates that the spatial dependence observed at lower bias currents disappears as the bias current exceeds the critical current, potentially indicating that only the weakest few locations are triggering in this regime. At high photon numbers and typical bias currents of 6.50$\mu$A, the device similarly operates outside the typical single-photon counting regime, wherein here the large number of simultaneous photons provide enough energy to drive large sections of the device normal at once. This increases the total hotspot resistance in the nanowire, and again eliminates the spatial dependence to the waveform as seen in the waveforms in Fig.~\ref{fig:highcases}b, where the waveforms now are uniform but have a much larger amplitude to the sinusoidal oscillations. LTSpice simulations confirm the increase in total hotspot resistance of the device, showing that increasing the hotspot resistance from 2 k$\Omega$ to 20 k$\Omega$ increases the amplitude of the position-dependent sinusoidal oscillations from 20\% of the voltage maximum to 50\% of the voltage maximum, consistent with the experimental waveforms observed in Fig.~\ref{fig:waveforms}a-b and Fig.~\ref{fig:highcases}b, respectively.  Simulations incorporating an electro-thermal model of the hotspot formation may better predict the high bias and high $\mu$ cases of operation where there is a combination of an increased number of hotspots, larger hotspot areas, and potential variations in the on time dynamics that cannot be captured with a simple switch. 

To date, most research involving SNSPDs has relied on single mode fiber-coupling or on complex multi-channel devices, and thus the spatial dependence of the SNSPD waveform within a single meander line has remained poorly explored.  As the demand for large-area SNSPDs increases for applications ranging from microscopy to free-space quantum communications, an exploration of spatial dependence like that presented here is increasingly critical to improving and optimizing device design. The experiments presented here show that large-area SNSPDs have a position dependent readout signal that varies between the coax connected end and grounded end of the detector due to pulse echos.  Additionally, this position dependence is evident not only in the bright count data, but also the dark count data, indicating that dark counts observed during typical single-photon counting operation arise from defect positions spanning the entire detector rather than from a small number of defect sites. Improvements in the position dependence reported here could lead to methods for filtering out known dark count signals in post processing or for spectrally resolved single photon detection for SNSPDs incorporating diffraction gratings.

\acknowledgments
This research was sponsored by the U. S. Department of Energy, Office of Science, Basic Energy Sciences, Materials Sciences and Engineering Division. Student (BEL) and postdoctoral (CEM) research support were provided by the IC Postdoctoral Research Fellowship Program at ORNL, administered by ORISE through an interagency agreement between the U.S. DOE and the Office of the Director of National Intelligence and by the DOE Science Undergraduate Laboratory Internships (SULI) program. SNSPD measurements with pulsed laser sources were conducted at the Center for Nanophase Materials Sciences, which is a DOE Office of Science User Facility.  The authors thank Matthew A. Feldman for support with microwave electronics.

The data that support the findings of this study are available from the corresponding author upon reasonable request.

%

\end{document}